\documentclass[twocolumn,showpacs,preprintnumbers,amsmath,amssymb,epsf]{revtex4-1}

\usepackage{graphicx,epsfig}
\usepackage{dcolumn}
\usepackage{bm}
\newcommand{\be}{\begin{equation}}
\newcommand{\ee}{\end{equation}}
\newcommand{\bse}{\begin{subequations}}
\newcommand{\ese}{\end{subequations}}
\newcommand{\bary}{\begin{eqnarray}}
\newcommand{\eary}{\end{eqnarray}}
\newcommand{\bwt}{\begin{widetext}}
\newcommand{\ewt}{\end{widetext}}

\begin{document}


\title{Possible High-Energy Neutrino and Photon Signals from
Gravitational Wave Bursts due to Double Neutron Star Mergers}
\author{He Gao$^{1}$, Bing Zhang$^{1,2}$,
Xue-Feng Wu$^{3}$ and Zi-Gao Dai$^{4}$}
\affiliation{
$^{1}$Department of Physics and Astronomy, University of Nevada, Las Vegas, NV 89154, USA, gaohe@physics.unlv.edu;zhang@physics.unlv.edu\\
$^{2}$Kavli Institute for Astronomy and Astrophysics and Department of Astronomy, Peking University, Beijing 100871, China\\
$^{3}$Purple Mountain Observatory, Chinese Academy of Sciences,
Nanjing 210008, China\\
$^{4}$School of Astronomy and Space Science,Nanjing University,
Nanjing 210093, China}


\begin{abstract}
As the technology of gravitational-wave and neutrino detectors
becomes increasingly mature, a multi-messenger era of astronomy is
ushered in. Advanced gravitational wave detectors are close to
making a ground-breaking discovery of gravitational wave bursts
(GWBs) associated with mergers of double neutron stars (NS-NS). It
is essential to study the possible electromagnetic (EM) and neutrino
emission counterparts of these GWBs. Recent observations and
numerical simulations suggest that at least a fraction of NS-NS
mergers may leave behind a massive millisecond magnetar as the
merger product. Here we show that protons accelerated in the forward
shock powered by a magnetar wind pushing the ejecta launched during
the merger process would interact with photons generated in the
dissipating magnetar wind and emit high energy neutrinos and
photons. We estimate the typical energy and fluence of the neutrinos
from such a scenario. We find that $\sim$PeV neutrinos could be
emitted from the shock front as long as the ejecta could be
accelerated to a relativistic speed.  The diffuse neutrino flux from
these events, even under the most optimistic scenarios, is too low
to account for the two events announced by the IceCube
Collaboration, but it is only slightly lower than the diffuse flux
of GRBs, making it an important candidate for the diffuse background
of $\sim$PeV neutrinos. The neutron-pion decay of these events make
them a moderate contributor to the sub-TeV gamma-ray diffuse
background.
\end{abstract}


 \maketitle

{\em I. Introduction.}The next-generation gravitational-wave (GW)
detectors, such as the advanced LIGO, advanced VIRGO and KAGRA
interferometers \cite{GWdetector}, are expected to detect GW signals
from mergers of two compact objects. One of the top candidates of
these gravitational wave bursts (GWBs) is the merger of two neutron
stars (i.e. NS-NS mergers) \cite{Merger}. The study of the
electromagnetic (EM) counterpart of such GWBs is of great interest
\cite{EMcontourpart}. Numerical simulations show that mergers of
binary neutron stars would leave two remnants, a postmerger compact
object and a mildly anisotropic ejecta with a typical velocity of
$\sim 0.1-0.3c$ (where $c$ is the speed of light) and typical mass
of $\sim 10^{-4}-10^{-2}M_{\odot}$ \cite{Ejecta}. Even though a
black hole is usually taken as the post-merger product,
observational data and numerical simulations suggest that for a
stiff equation of state of nuclear matter and a small enough total
mass of the two neutron stars, the postmerger product could be a
stable hypermassive, millisecond magneter
\citep{Dai06,Lattimer12,Zhang13,Gao13,giacomazzo13}. Recently, Ref.
\citep{Zhang13,Gao13} have systematically studied the EM signals for
the NS-NS scenario with a stable millisecond magnetar post-merger
product. Zhang \citep{Zhang13} proposed that the proto-magnetar
would eject a near-isotropic Poynting-flux-dominated outflow, the
dissipation of which would power a bright early X-ray afterglow for
essentially every GWB of NS-NS merger with a magnetar central
engine.  Gao et al. \citep{Gao13} proposed that after the
dissipation, within the framework of an energy injection scenario
\citep{dailu98}, a significant fraction of the wind energy would be
used to push the ejecta launched during the merger, which would
accelerate the ejecta to mildly or even highly relativistic speed,
making a strong external shock upon interaction with the ambient
medium.
Electrons are accelerated in
the shocked region, giving rise to broad band afterglow through
synchrotron emission \citep{Gao13}.

Protons are also expected to be accelerated in these shocks,
serving as efficient high-energy cosmic ray accelerators.
On the other hand, as propagating to us, photons emitted via
magnetic dissipation at a smaller radius from the engine
\citep{Zhang13} would first pass through
the external shock front, and have a good chance to interact with
the accelerated protons. Strong photo-meson interactions happen
at the $\Delta$-resonance, when the proton energy $E_p$ and
photon energy $E_\gamma$ satisfy the threshold condition
\begin{equation}
E_p E_\gamma \geq \frac{m_\Delta^2 - m_p^2}{2} \Gamma^2 = 0.147
~{\rm GeV}^2 \Gamma^2, \label{pgamma}
\end{equation}
where $\Gamma$ is the
bulk Lorentz factor,
$m_\Delta=1.232$ GeV and $m_p = 0.938$ GeV are the rest masses of
$\Delta^+$ and proton, respectively. The $\Delta^+$ particle decays
into two channels. The charged pion channel gives
$\Delta^+\rightarrow n\pi^+ \rightarrow n e^+ \nu_e \bar \nu_\mu
\nu_\mu$, with a typical neutrino energy $E_\nu \simeq 0.05 E_p$.
The neutron pion channel gives the $\Delta^+ \rightarrow p\pi^0
\rightarrow p\gamma\gamma$.

Note that the broad-band photons produced in the shocked region
could also serve as the seed photons for $p\gamma$ interaction.
However, since their peak flux in the X-ray band \citep{Gao13} is
much lower than that of the internal dissipation photons
\cite{Zhang13}, we do not consider their contribution.

With the multi-messenger era of astronomy ushered in, studying
multi-messenger signals in astrophysical sources is of the great
interest (e.g. \cite{Bartos11}). The high-energy neutrino detectors
such as IceCube have reached the sensitivity to detect high energy
neutrinos from astrophysical objects for the first time. Gamma-ray
bursts (GRBs) have been proposed to be one of the top candidates of
PeV neutrinos \cite{Waxman97}. However, a dedicated search of high
energy neutrinos coincident with GRBs have so far led to null
results \cite{Icecube1,Icecube2}, which already places a meaningful
constraint on GRB models \cite{Icecube2,He12,Zhang13b}. Very
recently, the IceCube collaboration announced their detections of
two neutrino events with an energy approximately 1-2
PeV\cite{Icecube13,Cholis12}, which could potentially represent the
first detections of high-energy neutrinos from astrophysical
sources. Among the proposed sources of such cosmic rays, GRBs stand
out as particularly capable of generating PeV neutrinos at this
level \cite{Cholis12,Liu13}. However, the absence of associated GRBs
for these two events calls for alternative cosmological PeV neutrino
sources. Here we investigate the possible neutrino signals
associated with NS-NS mergers with a millisecond magnetar central
engine using the photomeson interaction mechanism delineated above.

{\em II. General picture.} First of all, we adopt the {\em ansatz}
that NS-NS merger events leave behind a massive millisecond magnetar
and an essentially isotropic ejecta with mass $\sim (10^{-4}
-10^{-2}) {\rm M_\odot}$. Shortly after the merger, the neutron star
is able to cool down quickly so that a Poynting-flux-dominated
outflow can be launched \cite{Usov92,Metzger11}. Since the
postmerger magnetar would be initially rotating near the break-up
angular velocity, its total spin energy $E_{\rm{rot}}=(1/2)I
\Omega_{0}^{2} \simeq 2\times 10^{52} I_{45} P_{0,-3}^{-2} ~{\rm
ergs}$ (with $I_{45} \sim 1.5$ for a massive neutron star) may be
universal. Here $P_{0} \sim 1$ ms is the initial spin period of the
magnetar. Throughout the paper, the convention $Q=10^n Q_n$ is used
in cgs units, except for the ejecta mass $M_{\rm ej}$, which is in
units of solar mass $M_{\odot}$. Given nearly the same total energy,
the spindown luminosity and the characteristic spindown time scale
critically depend on the dipole magnetic field strength $B_{p}$,
i.e. $L_{\rm sd} = L_{\rm sd,0}/(1+t/T_{\rm sd})^2$, where $L_{\rm
sd,0} \simeq 10^{49} ~{\rm
erg~s^{-1}}~B^{2}_{p,15}R_{6}^{6}P_{0,-3}^{-4}$, and the spindown
time scale $T_{\rm{sd}} \simeq 2 \times 10^3 ~{\rm s}~I_{45}
B_{p,15}^{-2} P_{0,-3}^2 R_6^{-6}$, where $R=10^6R_6$ cm is the
stellar radius. Here we take the spindown luminosity $L_{\rm sd,0}$
as the total luminosity of the Poynting-flux-dominated outflow and
the spindown time scale $T_{\rm sd}$ as its duration. For
simplicity, we neglect the possible gravitational wave spin down of
the new-born magnetar \cite{Zhang01}. Note that both dipole magnetic
field strength and spindown timescale could have a relatively large
parameter space, which would add uncertainties to the following
results.

Initially, the heavy ejecta launched during the merger is not far
away from the magnetar, so that in a large solid angle range, the
magnetar wind would hit the ejecta before self-dissipation of the
magnetar wind happens. In this case, a good fraction ($\eta$) of the
magnetic energy may be rapidly discharged upon interaction between
the wind and the ejecta. The Thomson optical depth for a photon to
pass through the ejecta shell is $\tau_{\rm th} \sim \sigma_{\rm T}
M_{\rm ej} / (4\pi R^2 m_p)$. By setting the optical depth equals to
unity, we define a photosphere radius $R_{\rm
ph}=2.5\times10^{14}M_{\rm ej,-3}^{1/2}\,$cm for the ejecta. When
$R<R_{\rm ph}$, the spectrum of the dissipated wind is likely
quasi-thermal due to the large optical depth of photon scattering.
The typical photon energy can be estimated as $E_{\rm {ph,t}}\sim k
(L_{\rm sd,0}\eta/4\pi R^2 \sigma_{\rm SB})^{1/4}/\tau_{\rm th} \sim
27~{\rm eV}~L_{\rm sd,0,47}^{1/4}\eta_{-1}^{1/4}M_{\rm
ej,-4}^{-1}R_{14}^{3/2}$, where $\sigma_{\rm SB}$ is the
Stefan-Boltzmann constant. Alternatively, when $R>R_{\rm ph}$, the
typical synchrotron energy could be estimated as $E_{\gamma,t}
\simeq 1.8\times10^{4}~{\rm keV}~ L_{\rm sd,0,47}^{1/2} R_{15}^{-1}
\eta_{-1}^{3/2} \sigma_4^2$, where $\sigma$ is the magnetization
parameter of the Poynting flow when the magnetar wind catches the
ejecta \cite{Zhang11}.  In order to estimate the value of $\sigma$,
we assume that the proto-magnetar has $\sigma_0\sim 10^7$ at the
central engine and the magnetized flow is quickly accelerated to
$\Gamma \sim \sigma_0^{1/3}$ at $R_0 \sim 10^7$ cm, where $\sigma
\sim \sigma_0^{2/3}$ \cite{Komissarov09}. After this phase, the flow
may still accelerate as $\Gamma \propto R^{1/3}$, with $\sigma$
falling as $\propto R^{-1/3}$ \citep{Drenkhahn02}. Consequently, we
have $E_{\gamma,t}\simeq 1.8~{\rm keV}~ L_{\rm
sd,0,47}^{1/2}\eta_{-1}^{3/2}
\sigma_{0,7}^{4/3}R_{0,7}^{2/3}R_{15}^{-5/3}$.

As it is pushed forward by the magnetar wind, at a late time the
ejecta is far away enough from the central engine, so that before
hitting the ejecta, the magnetar wind already starts to undergo
strong self-dissipation, for instance, through
internal-collision-induced magnetic reconnection and turbulence
(ICMART) process \cite{Zhang11}. In this case, the typical
synchrotron frequency can be still estimated as above, except that
the emission radius is set to the self-dissipation radius, which we
parameterize as the ICMART radius $R_{\rm i}=10^{15}R_{\rm i,15}$,
rather than the blastwave radius \cite{Zhang11,Zhang13}, i.e.
$E_{\gamma,t} \simeq 1.8~{\rm keV}~ L_{\rm
sd,0,47}^{1/2}\eta_{-1}^{3/2} \sigma_{0,7}^{4/3}R_{0,7}^{2/3}R_{\rm
i,15}^{-5/3}$. Notice that for a substantial range of $M_{\rm ej}$,
we have $R_{\rm ph}<R_{\rm i}$. Overall, the seed photon energy for
$p\gamma$ interaction can be summarized as

\begin{eqnarray}
\label{Egamat} E_{\gamma,t} = \left\{ \begin{array}{ll} 27~{\rm
eV}~L_{\rm sd,0,47}^{1/4}\eta_{-1}^{1/4}M_{\rm
ej,-4}^{-1}R_{14}^{3/2}, ~~~~~~~ R<R_{\rm ph};\\
1.8~{\rm keV}~ L_{\rm sd,0,47}^{1/2}\eta_{-1}^{3/2}
\sigma_{0,7}^{4/3}R_{0,7}^{2/3}R_{15}^{-5/3},
R_{\rm ph}<R<R_{\rm i}; \\
1.8~{\rm keV}~ L_{\rm sd,0,47}^{1/2}\eta_{-1}^{3/2}
\sigma_{0,7}^{4/3}R_{0,7}^{2/3}R_{\rm i,15}^{-5/3},
R>R_{\rm i}; \\
\end{array} \right.
\end{eqnarray}

In the mean time, the magnetar-wind-powered ejecta would interact with
the ambient medium, forming a blastwave similar to GRB afterglow.
Depending on the unknown parameters such as $M_{\rm
ej}$, $B_p$ (and hence $L_{\rm sd,0}$) \cite{Gao13}, the blastwave
could be accelerated to a mildly or even highly relativistic speed,
due to the continuous energy injection from the magnetar wind.
Protons are accelerated from the forward shock front along with
electrons via the first-order Fermi acceleration process.
Consequently, when the seed photons due to magnetar wind
dissipation (Eq.\ref{Egamat}) pass through the shocked region,
significant neutrino production due to $p\gamma$ interaction
through $\Delta$-resonance would happen, as long as
the condition $\mathfrak{R}\equiv\frac{\Gamma\gamma_{\rm M} m_{\rm p}
c^2}{E_{\rm p,t}}>1$ is satisfied. Here, $E_{\rm p,t}=0.147 ~{\rm
GeV}^2 \Gamma^2/E_{\gamma,t}$ is the corresponding proton energy
for the typical seed photon at $\Delta$-resonance, and
$\gamma_{M}$ is the maximum proton Lorentz factor. It can be
estimated by balancing the acceleration time scale and the dynamical
time scale, which gives
$\gamma_{M}\sim \frac{\Gamma t e B'}{\zeta m_p c}$, where
$\zeta$ is a parameter of order unity that describes the details of
acceleration and $B'$ is the comoving magnetic field strength. Once
$p\gamma$ interaction happens, significant neutrinos with energy
$\epsilon_{\nu}\sim0.05E_{\rm p,t}$ would be released, the neutrino
emission fluence may be estimated as
\begin{eqnarray}
f_{\nu}=\frac{E_{\rm tot}\times f_{\gamma_{\rm p,t}} \times
f_{\pi}}{4\pi d^2},
\end{eqnarray}
where $E_{\rm tot}\sim4\pi R^3n\Gamma(\Gamma-1)m_{\rm p}c^2/3$ is
the total energy of all the protons, $f_{\gamma_{\rm
p,t}}\equiv\frac{E_{\gamma_{\rm p,t} }}{E_{\rm tot}}$ is the energy
fraction of the relevant protons, and $f_{\pi}$ is the fraction of
the proton energy that goes to pion production. Assuming a power-law
distribution of the shock accelerated protons: $N(E_p)dE_p \propto
E_{p}^{-p} dE_p$ (hereafter assuming $p>2$), one can obtain
$f_{\gamma_{\rm p,t}}=\left(\frac{\gamma_{\rm
p,t}}{\gamma_m}\right)^{2-p}$, where
$\gamma_m=(\Gamma-1)\frac{p-2}{p-1}+1$ is the minimum proton Lorentz
factor. The fraction of the proton energy that goes to pion
production could be estimated as $f_{\pi}\equiv\frac{1}{2}(
1-(1-<\chi_{p\rightarrow \pi}>)^{\tau_{p\gamma}})$, where
$\tau_{p\gamma}$ is the $p\gamma$ optical depth and
$<\chi_{p\rightarrow\pi}> \simeq 0.2$ is the average fraction of
energy transferred to pion. Notice that $f_{\pi}$ is roughly
proportional to $\tau_{p\gamma}$ when $\tau_{p\gamma} < 3$
\cite{Zhang13b}.

{\em III. Neutrino energy and fluence.} The dynamics of the
blastwave is defined by energy conservation \cite{Gao13}
\begin{eqnarray}\label{Dyn}
L_{\rm{0}}t=(\gamma-1)M_{\rm ej}c^2+(\gamma^{2}-1)M_{\rm sw}c^2,
\end{eqnarray}
where $L_0=\xi L_{\rm sd,0}$ is the magnetar injection luminosity
into the blastwave,
and $M_{\rm sw}=(4\pi/3)R^3nm_p$ is the swept-up mass from the
interstellar medium. Initially, one has $(\gamma-1)M_{\rm
ej}c^2\gg(\gamma^{2}-1)M_{\rm sw}c^2$, so that the kinetic energy of the
ejecta would increase linearly with time until $R={\rm min}(R_{\rm
sd}, R_{\rm dec})$, where the deceleration radius $R_{\rm{dec}}$ is
defined by the condition $(\gamma-1)M_{\rm
ej}c^2=(\gamma^{2}-1)M_{\rm sw}c^2$. By setting $R_{\rm{dec}}\sim
R_{\rm sd}$, we can derive a critical ejecta mass
\begin{eqnarray}
M_{\rm ej,c,1}\sim 10^{-3}
M_{\odot}n^{1/8}I_{45}^{5/4}B^{-3/4}_{p,14}R_{6}^{-9/4}P_{0,-3}^{-1}
\xi^{7/8},
\end{eqnarray}
which separate regimes with different blastwave dynamics \cite{Gao13}:

Case I: $M_{\rm ej} < M_{\rm ej,c,1}$ or $R_{\rm sd}>R_{\rm dec}$.
In such case, the ejecta can be accelerated linearly until the
deceleration radius $R_{\rm dec}\sim3.9\times10^{17}M_{\rm
ej,-4}^{2/5}L_{\rm sd,0,47}^{-1/10}n_0^{-3/10}$, where bulk Lorentz
factor of the blastwave is $\Gamma_{\rm{dec}}\sim12.2L_{\rm
sd,0,47}^{3/10}M_{\rm ej,-4}^{-1/5}n_0^{-1/10}$. After that, the
blastwave decelerates, but still with continuous energy injection
until $R_{\rm sd}\sim1.0\times10^{18}\xi^{1/2}L_{\rm
sd,0,47}^{-1/4}n_0^{-1/4}$, where
$\Gamma_{\rm{sd}}\sim7.5\xi^{-1/4}L_{\rm sd,0,47}^{3/8}n_0^{-1/8}$.
During the acceleration phase, the blastwave passes the
non-relativistic to relativistic transition line $\Gamma-1=1$ at
radius $R_{\rm N}\sim2.2\times10^{14}M_{\rm ej,-4}L_{\rm
sd,0,47}^{-1}$.

For the different radius range of the typical photon energy shown in
Eq. \ref{Egamat}, we can investigate whether $p\gamma$ interaction
at $\Delta$-resonance can occur, and if so, the typical energy and
fluence of neutrino emission. We first assume that the blastwave is
always non-relativistic when $R<=R_{\rm ph}$, since $R_{\rm N}$ is
comparable with $R_{\rm ph}$ with a high probability. In this range,
we have $\mathfrak{R}=0.1 \eta_{-1}^{1/4}L_{\rm
sd,0,47}^{-5/12}M_{\rm ej,-4}^{-1/3}n_0^{1/2} R_{14}^{11/6} < 1$,
implying that p$\gamma$ interaction at $\Delta$-resonance could
hardly happen. Second, at $R_{\rm ph}<R<R_{\rm i}$, we have
$\mathfrak{R}=26.0 \eta_{-1}^{3/2}L_{\rm sd,0,47}^{-1/6}M_{\rm
ej,-4}^{2/3}n_0^{1/2} \sigma_{0,7}^{4/3}R_{0,7}^{2/3} R_{15}^{-4/3}
> 1$, so that $p\gamma$ interaction would happen at
$\Delta$-resonance. The typical neutrino energy and fluence could be
estimated as $\epsilon_{\nu}=1.1\times10^{-2}~\rm{PeV}~
\eta_{-1}^{-3/2}L_{\rm sd,0,47}^{1/6}M_{\rm
ej,-4}^{-2/3}\sigma_{0,7}^{-4/3}R_{0,7}^{-2/3} R_{15}^{7/3}$, and
$f_{\nu}=1.6\times10^{-12} \eta_{-1}^{-0.05}L_{\rm
sd,0,47}^{0.65}n_0^{}\sigma_{0,7}^{-0.93}R_{0,7}^{-0.47}
R_{15}^{3.2}$. Next, similar to the previous stage, at $R_{\rm
i}<R<R_{\rm dec}$, we have $\mathfrak{R}=120.7 \eta_{-1}^{3/2}L_{\rm
sd,0,47}^{-1/6}M_{\rm ej,-4}^{2/3}n_0^{1/2}
\sigma_{0,7}^{4/3}R_{0,7}^{2/3}R_{\rm i,15}^{-5/3}R_{17}^{1/3} > 1$,
and the typical neutrino energy and fluence are
$\epsilon_{\nu}=0.21~\rm{PeV}~ \eta_{-1}^{-3/2}L_{\rm
sd,0,47}^{1/6}M_{\rm
ej,-4}^{-2/3}\sigma_{0,7}^{-4/3}R_{0,7}^{-2/3}R_{\rm i,15}^{5/3}
R_{17}^{2/3}$ and $f_{\nu}=1.6\times10^{-8} \eta_{-1}^{-0.05}L_{\rm
sd,0,47}^{0.65}n_0^{}\sigma_{0,7}^{-0.93}R_{0,7}^{-0.47}R_{\rm
i,15}^{1.17} R_{17}^{2}$, respectively. Finally, when approaching
the spin-down radius, i.e., $R_{\rm dec}<R<R_{\rm sd}$, one has
$\mathfrak{R}=1.2\times10^{3}
\eta_{-1}^{3/2}n_0^{}\sigma_{0,7}^{4/3}R_{0,7}^{2/3}R_{\rm
i,15}^{-5/3} R_{18}^{2} > 1$, and the typical neutrino energy and
fluence are $\epsilon_{\nu}=0.24~\rm{PeV}~
\eta_{-1}^{-3/2}n_0^{-1/2}\sigma_{0,7}^{-4/3}R_{0,7}^{-2/3}R_{\rm
i,15}^{5/3} R_{18}^{-1}$ and $f_{\nu}=1.6\times10^{-6}
\eta_{-1}^{-0.05}L_{\rm
sd,0,47}^{0.65}n_0^{}\sigma_{0,7}^{-0.93}R_{0,7}^{-0.47}R_{\rm
i,15}^{1.17} R_{18}^{2}$, respectively. For better illustration, we
take $L_{sd,0} = 10^{47}$ and $M_{\rm ej}= 10^{-4} M_{\odot}$ as an
example and plot the evolution of $\epsilon_{\nu}$ and $f_{\nu}$ for
this dynamical case in Figure 1.

Case II: $M_{\rm ej} \sim M_{\rm ej,c,1}$ or $R_{\rm sd} \sim R_{\rm
dec}$. In this case, the ejecta would be continuously accelerated
until $R_{\rm sd}=1.2\times10^{18}\xi^{3}L_{\rm sd,0,49}^{-1}M_{\rm
ej,-4}^{-2}$, where the bulk Lorentz factor reaches $\Gamma_{\rm
sd}=83.3\xi^{}M_{\rm ej,-4}^{-1}$. Similar to case I, for $R<=R_{\rm
ph}$, we do not expect significant $p\gamma$ interaction since
$\mathfrak{R}=0.01 \eta_{-1}^{1/4}L_{\rm sd,0,49}^{-5/12}M_{\rm
ej,-4}^{-1/3}n_0^{1/2} R_{14}^{11/6} < 1$. In the next stage $R_{\rm
ph}<R<R_{\rm i}$, one has $\mathfrak{R}=12.0 \eta_{-1}^{3/2}L_{\rm
sd,0,49}^{-1/6}M_{\rm
ej,-4}^{2/3}n_0^{1/2}\sigma_{0,7}^{4/3}R_{0,7}^{2/3}
R_{15}^{-4/3}>1$. The expected neutrino energy and fluence are
$\epsilon_{\nu}=0.02~\rm{PeV}~ \eta_{-1}^{-3/2}L_{\rm
sd,0,49}^{1/6}M_{\rm ej,-4}^{-2/3}\sigma_{0,7}^{-4/3}R_{0,7}^{-2/3}
R_{15}^{7/3}$, and $f_{\nu}=3.2\times10^{-11}
\eta_{-1}^{-0.05}L_{\rm
sd,0,49}^{0.65}n_0^{}\sigma_{0,7}^{-0.93}R_{0,7}^{-0.47}
R_{15}^{3.2}$, respectively. Finally, at $R_{\rm i}<R<R_{\rm sd}$,
one has $\mathfrak{R}=55.9\eta_{-1}^{3/2}L_{\rm
sd,0,49}^{-1/6}M_{\rm
ej,-4}^{2/3}n_0^{1/2}\sigma_{0,7}^{4/3}R_{0,7}^{2/3}R_{\rm
i,15}^{-5/3} R_{17}^{1/3} > 1$, and $\epsilon_{\nu}=0.5~\rm{PeV}~
\eta_{-1}^{-3/2}L_{\rm sd,0,49}^{1/6}M_{\rm
ej,-4}^{-2/3}\sigma_{0,7}^{-4/3}R_{0,7}^{-2/3}R_{\rm i,15}^{5/3}
R_{17}^{2/3}$, $f_{\nu}=3.2\times10^{-7}\eta_{-1}^{-0.05}L_{\rm
sd,0,49}^{0.65}n_0^{}\sigma_{0,7}^{-0.93}R_{0,7}^{-0.47}R_{\rm
i,15}^{1.17} R_{17}^{2}$, respectively. In this case, we take
$L_{sd,0} = 10^{49}$ and $M_{\rm ej}= 10^{-4} M_{\odot}$, and plot
the evolution of $\epsilon_{\nu}$ and $f_{\nu}$ in Figure 1.

Case III: $M_{\rm ej} > M_{\rm ej,c,1}$ or $R_{\rm sd} < R_{\rm
dec}$. Similar to Case II, the ejecta would be accelerated to
a relativistic speed of $\Gamma_{\rm sd}=16.7\xi M_{\rm ej,-3}^{-1}$
until $R_{\rm sd}=5.0\times10^{16}\xi^{3}L_{\rm sd,0,49}^{-1}
M_{\rm ej,-3}^{-2}$. Similarly, when $R \leq R_{\rm ph}$, one has
$\mathfrak{R}=0.004\eta_{-1}^{1/4}L_{\rm sd,0,49}^{-5/12}M_{\rm
ej,-3}^{-1/3}n_0^{1/2} R_{14}^{11/6} < 1$, and hence, no significant
neutrino emission. At $R_{\rm ph}<R<R_{\rm i}$, one has
$\mathfrak{R}=35.1\eta_{-1}^{3/2}L_{\rm sd,0,49}^{-1/6}M_{\rm
ej,-3}^{2/3}n_0^{1/2}\sigma_{0,7}^{4/3}R_{0,7}^{2/3}
R_{15}^{-4/3}>1$, and $\epsilon_{\nu}=8.4\times10^{-3}~\rm{PeV}~
\eta_{-1}^{-3/2}L_{\rm sd,0,49}^{1/6}M_{\rm
ej,-3}^{-2/3}\sigma_{0,7}^{-4/3}R_{0,7}^{-2/3} R_{15}^{7/3}$,
$f_{\nu}=3.2\times10^{-11} \eta_{-1}^{-0.05}L_{\rm
sd,0,49}^{0.65}n_0^{}\sigma_{0,7}^{-0.93}R_{0,7}^{-0.47}
R_{15}^{3.2}$. At $R_{\rm i}<R<R_{\rm sd}$, one has
$\mathfrak{R}=163.1 \eta_{-1}^{3/2}L_{\rm sd,0,49}^{-1/6}M_{\rm
ej,-3}^{2/3}n_0^{1/2}\sigma_{0,7}^{4/3}R_{0,7}^{2/3}R_{\rm
i,15}^{-5/3} R_{17}^{1/3} > 1$, and $\epsilon_{\nu}=0.2~\rm{PeV}~
\eta_{-1}^{-3/2}L_{\rm sd,0,49}^{1/6}M_{\rm
ej,-3}^{-2/3}\sigma_{0,7}^{-4/3}R_{0,7}^{-2/3}R_{\rm i,15}^{5/3}
R_{17}^{2/3}$, $f_{\nu}=3.2\times10^{-7}\eta_{-1}^{-0.05}L_{\rm
sd,0,49}^{0.65}n_0^{}\sigma_{0,7}^{-0.93}R_{0,7}^{-0.47}R_{\rm
i,15}^{1.17} R_{17}^{2}$. For this case, we take $L_{sd,0} =
10^{47}$ and $M_{\rm ej}= 10^{-3} M_{\odot}$ and plot the evolution
of $\epsilon_{\nu}$ and $f_{\nu}$ in Figure 1.

Note that there is another critical ejecta mass $M_{\rm ej,c,2}\sim
6\times 10^{-3} M_\odot I_{45} P_{0,-3}^{-2} \xi$ (defined
by setting $E_{\rm rot}\xi = 2 (\gamma-1) M_{\rm ej,c,2} c^2$),
above which the blast wave would never reach a relativistic speed
\cite{Gao13}. The dynamics is similar to
Case III, with the coasting regime in the non-relativistic phase. In
this case, we always have $\mathfrak{R}<1$, therefore no significant
neutrino flux is expected.

\begin{figure}[t!]
\vspace{0.3cm} {\centering
\resizebox*{0.5\textwidth}{0.3\textheight}
{\includegraphics{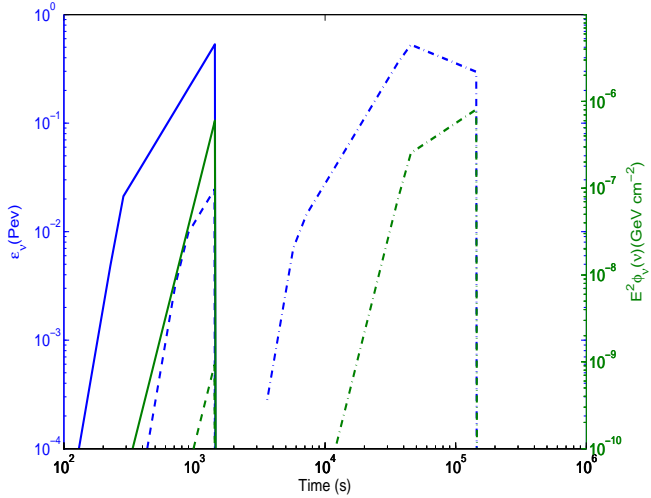}}
\par}
\caption{Examples of the evolution of neutrino energy
$\epsilon_{\nu}$ and fluence $f_{\nu}$ for different dynamics: Case
I (dash-dot), Case II (solid) and Case III (dashed). Blue lines
represent $\epsilon_{\nu}$ and green lines show $f_{\nu}$. Model
parameters: $n_0=1$,$\eta= 0.1$, $\sigma_{0}=10^7$, $R_{0}=10^7$,
and $D=300$ Mpc (the advanced LIGO horizon for NS-NS mergers). For
the magnetar parameters for each case, see text. } \label{Fig1}
\end{figure}

{\em IV. Detection prospect.} From the above calculation, one can
see that when the post-merger product is a millisecond magnetar and
the outgoing ejecta could be accelerated to a relativistic speed,
$\sim$PeV neutrinos could indeed be emitted from NS-NS mergers
scenario. These neutrinos are well suited for detection with
IceCube\cite{Ahrens04}.

As shown in Figure 1, for different initial conditions, i.e.,
different combinations of $M_{\rm ej}$ and $L_{\rm sd}$, the maximum
neutrino fluence is always reached at the spin-down time scale. We
therefore take the neutrino energy and fluence at this epoch as the
typical values for each specific NS-NS merger event. For the events
happening at 300 Mpc, the optimistical typical neutrino fluence
could be as large as $10^{-6}- 10^{-5}\rm{GeV~ cm^{-2}}$
(corresponding to $\sigma_0=10^{7},~10^{6}$ respectively), one or
two orders of magnitude lower than the typical fluence of
GRBs\citep{He12}. Given the typical neutrino energy $\sim$PeV and
the IceCube effective area $\sim$ several $10^6~{\rm cm^2}$
\cite{Ahrens04,Li12}, optimistically only several $10^{-6} -
10^{-5}$ neutrinos are expected to be detected by IceCube for a
single event.

In any case, these events would contribute to the $\sim$PeV neutrino
background. The NS-NS merger event rate is rather uncertain, i.e.,
$(10 - 5\times 10^{4}) ~{\rm Gpc^{-3}~yr^{-1}}$ \citep{EventRate}.
Considering that only a fraction of NS-NS merger event would leave
behind a massive neutron star rather than a black hole, and that
only a sub-fraction of these mergers have the right $M_{\rm ej}$ and
$L_{\rm sd,0}$ to make relativistic blastwaves, the event rate of
NS-NS mergers that generate PeV neutrinos may be at least one order
of magnitude lower, i.e. $\sim (1-5\times 10^3) ~{\rm
Gpc^{-3}~yr^{-1}}$. Even with the most optimistic estimate, the
$\sim$PeV diffuse back ground is $\sim 10^{-10}~{\rm
GeV~cm^{-2}~s^{-1}~sr^{-1}}$. It takes tens of years to get two
events. So these systems are not likely the origin of the two
reported PeV events announced by the Icecube collaboration
\cite{Icecube13}. Nevertheless, compared with the GRB event rate
$1~{\rm Gpc^{-3}~yr^{-1}}$ \cite{GRBeventrate}, this scenario may
gain the event rate by 1-2 orders of magnitude than GRBs. Noticing
that a typical GRB has a fluence 1-2 orders of magnitude higher than
a magnetar-wind-powered NS-NS merger remnant, our scenario could
contribute to the $\sim$PeV neutrino diffuse background, which is
comparable or slightly lower than that of GRBs.

{\em V. High energy photon emission.} Besides high-energy neutrino
emission, the decay of $\pi^0$ produced in $p\gamma$ interactions
would lead to the production of high energy gamma-ray photons.
Assuming that half of the $\Delta^+$ decays go to the $\pi^+$
channel (neutrino production), while the other half go to the
$\pi^0$ channel ($\gamma$-ray production), the typical gamma-ray
photon energy and fluence values would be comparable to the
neutrinos we studied in section \emph{III}. However, such
high-energy photons may interact with the synchrotron emission
photons in the shock \citep{Gao13} to produce electron/positron
pairs, $\gamma\gamma\rightarrow e^{\pm}$, and initiate an
electromagnetic cascade: the pairs would emit photons via
synchrotron and inverse Compton, which would be converted back to
pairs, and the pairs would emit photons again, etc. Photons can
escape only when the $\gamma\gamma$ optical depth becomes lower than
unity \cite{Murase}. Following the calculation shown in
Ref.\cite{Gao13}, we find that the $\gamma\gamma$ optical depth
exceeds unity for photon energy above $\epsilon_{\gamma\gamma}\sim
100 \rm{GeV}$. For simplicity, we assume that the total energy of
the $\pi^0$-decay photons would finally show up around $100
\rm{GeV}$ through an EM cascade. These photons are within the energy
windows of the Fermi/LAT. In the most optimistic situation, the
photon flux for an event at 300 Mpc could be as high as $10^{-11}\rm
erg~ cm^{-2}~ s^{-1}$, which is essentially $10^{-10}\rm~ photons~
cm^{-2}~ s^{-1}$. The effective area of LAT for $100 \rm{GeV}$
photons is around 9000 $\rm cm^2$ \cite{Atwood09}, suggesting that
even for $T_{\rm sd}\sim 10^{5}$, one single NS-NS merger event
could not trigger LAT. Nevertheless, the total diffuse flux from
these events could reach $\sim$ several $10^{-7}~{\rm
MeV~cm^{-2}~s^{-1}~sr^{-1}}$ optimistically, giving a moderate
contribution to the sub-TeV $\gamma$-ray background, i.e.,
$4\times10^{-4}~{\rm MeV~cm^{-2}~s^{-1}~sr^{-1}}$, according to
Fermi/LAT observation\cite{Abdo10}.

{\em VI. Acknowledge.} We thank stimulative discussions with Zhuo Li
and Qiang Yuan. We acknowledge the National Basic Research Program
(``973" Program) of China (Grant No. 2009CB824800 and 2013CB834900),
National Science Foundation (AST-0908362), and National Natural
Science Foundation of China (grant No. 11033002 \& 10921063). XFW
acknowledges support by the One-Hundred-Talents Program and the
Youth Innovation Promotion Association of Chinese Academy of
Sciences.

\end{document}